\documentstyle[12pt]{article}

\begin{document}

\begin{titlepage}
\vskip 2cm
\begin{center}
{\Large\bf A magnetization equation for non-equilibrium spin systems
} \vskip 3cm {\bf Fuad M. Saradzhev\footnote{{\tt E-mail:
fsarajov@phys.ualberta.ca} }
\\} \vskip 5pt
{\sl Department of Physics, University of Alberta\\
 Edmonton, Alberta, Canada T6G 2G7\\} \vskip 2pt

\end{center}
\vskip .5cm \rm

\begin{abstract}
A magnetization equation for a system of spins evolving
non-adiabatically and out of equilibrium is derived without
specifying the internal interactions. For relaxation processes, this
equation provides a general form of magnetization damping. A special
case of the spin-spin exchange interaction is considered.

\end{abstract}

\vskip .5cm \rm

Pacs numbers: 76.20.+q, 72.25.Ba.

\end{titlepage}






\section{Introduction}

Magnetization dynamics is conventionally described by the
Landau-Lifshitz-Gilbert (LLG) equation \cite{dau},\cite{gil}, which
provides a plausible phenomenological model for many experimental
results. Recently, the LLG equation and the Gilbert damping term
have been derived from an effective Hamiltonian including the
radiation-spin interaction (RSI) \cite{ho}. It has been assumed
there that the spin system maintains quasi-adiabatic evolution.

Various kinds of relaxation processes are usually melded together
into a single damping term. Relativistic relaxation processes result
in the Gilbert damping term with one damping parameter, while for
the case of both exchange and relativistic relaxation the damping
term is a tensor with several damping parameters \cite{bar},
\cite{saf}.

The relaxation processes are specified by interactions of spins with
each other and with other constituents of the magnetic system. A
derivation of the damping term from first principles should
therefore start with a microscopic description of the interactions.
Even though such microscopic derivation of damping has been
performed for some relaxation processes (for instance, \cite{kam}),
a full version of derivation for the Gilbert damping term has not
yet been given, and in particular for a system in non-equilibrium.

In the present talk, we aim to derive a magnetization equation for a
general non-equilibrium spin system without specifying the
interaction Hamiltonian and related relaxation processes. We start
with a system of spins precessing in the effective magnetic field
$\vec{\bf H}_{eff}$ neglecting for a moment mutual interactions.
Then at a fixed time later interactions in the system are switched
on and influence the original precessional motion.

The interactions are assumed to be time-dependent, and the spin
system evolves non-adiabatically out of equilibrium trying to relax
to a new equilibrium magnetization. We perform a transformation,
which is analogous to the one used in the transition to the
interaction picture, to connect the density and magnetic moment
operators before and after the time, when a new non-equilibrium
dynamics starts, and to find an explicit expression for the
interaction contribution to the magnetization equation.

\section{Magnetization Equation}

Let us consider a quantum spin system defined by
\begin{equation}
\hat{\cal H} = \hat{\cal H}_0 + {\lambda} \hat{\cal H}_{I},
\label{1}
\end{equation}
where $\hat{\cal H}_0$ is the Zeeman Hamiltonian describing the
interaction of spins with an effective magnetic field
\begin{equation}
\hat{\cal H}_0 = - {\gamma} \sum_{i} \hat{S}_{i} \cdot \vec{\bf
H}_{eff}(t), \label{2}
\end{equation}
${\gamma}$ being the gyromagnetic ratio and $\hat{S}_i$ being the
spin operator of the ith atom. The effective magnetic field
$\vec{\bf H}_{eff}$ is given by the energy variational with
magnetization , $\vec{\bf H}_{eff}=- {\delta}E(M)/{\delta}\vec{\bf
M}$, where $E(M)$ is the free energy of the magnetic system. This
field includes the exchange field, the anisotropy field, and the
demagnetizing field, as well as the external field, $\vec{\bf
H}_{ext}$.

The Hamiltonian $\hat{\cal H}_{I}$ represents other possible types
of interactions, which can include, for instance, higher order
spin-spin interactions. The interaction terms included in $\hat{\cal
H}_{I}$ are in general time-dependent, being switched on
adiabatically or instantly at a fixed time $t_0$. The parameter
${\lambda}$ in (\ref{1}) can be chosen small in order to take into
account the higher order effects perturbatively.

We introduce next the magnetic moment operator
\begin{equation}
\hat{\cal M} \equiv - \frac{{\delta}\hat{\cal H}}{{\delta}\vec{\bf
H}_{ext}}, \label{3}
\end{equation}
which is the response of the spin system to the external field. The
magnetization is defined as an ensemble average of the response
\begin{equation}
\vec{\bf M}=\langle \hat{\cal M} \rangle \equiv \frac{1}{V} {\bf \rm
Tr}\{ \hat{\rho} \hat{\cal M} \}, \label{4}
\end{equation}
where $V$ is the volume of the system, while $\hat{\rho}$ is the density operator satisfying the quantum
Liouville-von Neumann (LvN) equation
\begin{equation}
i \hbar \frac{{\partial}\hat{\rho}}{{\partial}t} + [ \hat{\rho},
\hat{\cal H} {]}_{-}=0. \label{llg}
\end{equation}
For systems in equilibrium,
the Hamiltonian itself satisfies the LvN equation and the density
operator is expressed in terms of the Hamiltonian. For
non-equilibrium systems, the density operator is constructed by
making use of the time-dependent adiabatic invariants
\cite{lewis},\cite{kim}.

To derive the magnetization equation, we proceed as follows. We
perform, on $\hat{\rho}(t)$, the transformation
\begin{equation}
\hat{\rho} \to \hat{\rho}_{int} \equiv \hat{U}(t_0,t) \hat{\rho}(t)
\hat{U}(t,t_0) \label{13}
\end{equation}
defined by the operator
\begin{equation}
\hat{U}(t_0,t) \equiv T\exp \{ \frac{i}{\hbar} \int_{t_0}^{t}
d{\tau} \hat{\cal H}_0({\tau}) \} \label{11} ,
\end{equation}
where $T$ denotes the time-ordering operator. For systems with
$\hat{\cal H}_0$ constant in time, the operator $\hat{U}(t_0,t)=
\exp\{ (i/{\hbar}) \hat{\cal H}_0 (t-t_0) \}$ leads to the
interaction picture, which proves to be very useful for all forms of
interactions since it distinguishes among the interaction times. For
our system with both $\hat{\cal H}_0$ and $\hat{\cal H}_{I}$
dependent on time, the operator (\ref{11}) plays the same role,
removing the unperturbed part of the Hamiltonian from the LvN
equation.

Substituting Eq.(\ref{13}) into (\ref{llg}), yields
\begin{equation}
i{\hbar} \frac{{\partial}\hat{\rho}_{int}}{{\partial}t} = {\lambda}
\Big[ \hat{\cal H}_{int}, \hat{\rho}_{int} {\Big]}_{-}, \label{14}
\end{equation}
where
\begin{equation}
\hat{\cal H}_{int}(t) \equiv \hat{U}(t_0,t) \hat{\cal H}_{I}(t)
\hat{U}(t,t_0). \label{15}
\end{equation}
The magnetic moment operator and the magnetization become
\begin{equation}
\hat{\cal M}=\hat{\cal M}_0 + \hat{\cal M}_{I} \label{16}
\end{equation}
and
\begin{equation}
\vec{\bf M} = \frac{1}{V} {\bf \rm Tr} \{ \hat{\rho}_{int} (
\hat{\cal M}_{0,int} + \hat{\cal M}_{I,int} ) \}, \label{18}
\end{equation}
where
\begin{equation}
\hat{\cal M}_0 = - \frac{{\delta}\hat{\cal H}_0}{{\delta}\vec{\bf
H}_{ext}} = {\gamma} \sum_{i} \hat{S}_i \label{6}
\end{equation}
and
\begin{equation}
\hat{\cal M}_{I} \equiv - {\lambda} \frac{{\delta} \hat{\cal
H}_{I}}{{\delta} \vec{\bf H}_{ext}}, \label{17}
\end{equation}
while $\hat{\cal M}_{0,int}$ and $\hat{\cal M}_{I,int}$ are related
with $\hat{\cal M}_0$ and $\hat{\cal M}_{I}$, respectively, in the
same way as $\hat{\cal H}_{int}$ is related with $\hat{\cal H}_{I}$.
The operators $\hat{\cal M}_{0}^{a}$, $a=1,2,3$, fulfill the $SU(2)$
magnetization algebra
\begin{equation}
\Big[ \hat{\cal M}_{0}^{a} , \hat{\cal M}_{0}^{b} {\Big]}_{-} =
i{\hbar}{\gamma} {\varepsilon}^{abc} \hat{\cal M}_{0}^{c}, \label{8}
\end{equation}
where the summation over repeated indices is assumed.

The operators $\hat{\cal M}_{0,{int}}$, $\hat{\cal M}_{I,{int}}$ are
generally used to calculate the magnetic susceptibility
\cite{white}. Let us show now how these operators determine the time
evolution of magnetization. The evolution in time of $\hat{\cal
M}_{0,{int}}$ is given by the equation
\[
\frac{{\partial}\hat{\cal M}_{0,{int}}}{{\partial}t} =
\frac{i}{\hbar} \hat{U}(t_0,t) \Big[ \hat{\cal H}_0, \hat{\cal M}_0
{\Big]}_{-} \hat{U}(t,t_0)
\]
\begin{equation}
= {\gamma} \hat{\cal M}_{0,{int}} \times \vec{\bf H}_{eff}.
\label{19}
\end{equation}
It describes the magnetization precessional motion with respect to
$\vec{\bf H}_{eff}$.

The equation for $\hat{\cal M}_{I,{int}}$ describes more complex
magnetization dynamics governed by the interaction Hamiltonian
$\hat{\cal H}_{I}$. However, this dynamics includes the precessional
motion as well. Introducing
\begin{equation}
\vec{\bf D}_{I} \equiv \frac{i}{\hbar} \Big[ \hat{\cal H}_0 ,
\hat{\cal M}_{I} {\Big]}_{-} - {\gamma} \hat{\cal M}_{I} \times
\vec{\bf H}_{eff} \label{20}
\end{equation}
to represent deviations from the purely precessional motion, we
bring the equation for $\hat{\cal M}_{I,{int}}$ into the following
form
\[
\frac{{\partial}\hat{\cal M}_{I,{int}}}{{\partial}t} = {\gamma}
\hat{\cal M}_{I,{int}} \times \vec{\bf H}_{eff}
\]
\begin{equation}
+ \hat{U}(t_0,t) \Big( \frac{{\partial}\hat{\cal
M}_{I}}{{\partial}t} + \vec{\bf D}_{I} \Big) \hat{U}(t,t_0).
\label{21}
\end{equation}
Taking the time-derivative of Eq.(\ref{18}) and using
Eqs.(\ref{14}),(\ref{19}) and (\ref{21}), we finally obtain
\begin{equation}
\frac{d\vec{\bf M}}{dt} = - |{\gamma}| \vec{\bf M} \times \vec{\bf
H}_{eff} + \vec{\bf D}, \label{22}
\end{equation}
where
\begin{equation}
\vec{\bf D} \equiv {\lambda} \langle \frac{1}{i{\hbar}} \Big[
\hat{\cal M}, \hat{\cal H}_{I} {\Big]}_{-} \rangle + \langle
\frac{{\partial}\hat{\cal M}_{I}}{{\partial}t} + \vec{\bf D}_{I}
\rangle. \label{23}
\end{equation}
Therefore, Eq.(\ref{22}) is the magnetization equation for the
system specified by (\ref{1}). This equation is general since it is
derived without specifying $\hat{\cal H}_{I}$. The $\vec{\bf
D}$-term contains all effects that the interactions, $\hat{\cal
H}_{I}$, can have on the magnetization precession, so that
Eq.(\ref{22}) is complete.

The contribution of $\hat{\cal H}_{I}$ to the $\vec{\bf D}$-term in
the magnetization equation can be divided into two parts. One is
proportional to $\langle [ \hat{\cal M} , \hat{\cal H}_{I} {]}_{-}
\rangle$ and is related to the change in the density matrix when the
interactions of $\hat{\cal H}_{I}$ are switched on. The second part
$\langle \frac{{\partial} \hat{\cal M}_{I}}{{\partial}t} + \vec{\bf
D}_{I} \rangle$ originates from the change in the magnetization
itself. Which part of $\vec{\bf D}$ is dominating depends on the
nature of the interactions. For the interactions related to the
relaxation processes, the $\vec{\bf D}$-term represents a general
form of magnetization damping.

\section{Example: spin-spin interactions}

The spin-spin interactions among the spins in the system introduce
many body effects, which can be treated perturbatively in the weak
coupling regime. In this case the $\vec{\bf D}$-term can be expanded
in powers of $\lambda$. To demonstrate this, we consider the
spin-spin interactions of a specific type. The interaction between
spins is usually an exchange interaction of the form
\begin{equation}
-2J \sum_{i,j} \hat{S}_i \hat{S}_j = - \frac{2J}{{\gamma}^2}
\hat{\cal M}_0^2, \label{52}
\end{equation}
the coupling constant $J$ being called the exchange integral. We
generalize the ansatz given by Eq.(\ref{52}) by assuming that the
exchange integral depends on the magnetization and introduce the
spin-spin interactions as follows
\begin{equation}
{\lambda}\hat{\cal H}_{I} = \sum_{i,j} J^{ab}(M) \hat{S}_i^a
\hat{S}_j^b, \label{53}
\end{equation}
where $J^{ab}={\lambda}M^aM^b$. Since $\hat{\cal H}_{I}$ does not
depend explicitly on the external field, its contribution to the
magnetic moment operator vanishes, $\hat{\cal M}_{I}=0$.

The non-vanishing commutator $[\hat{\cal M}_0, \hat{\cal H}_{I}
{]}_{-}$ in Eq.(\ref{23}) is the only contribution of the spin-spin
interaction to the magnetization equation, resulting in
\begin{equation}
\vec{\bf D} = \frac{\lambda}{\gamma} \vec{\bf M} \times
\vec{\bf{\Omega}}, \label{54}
\end{equation}
where
\begin{equation}
{\Omega}^a \equiv \langle \Big[ \hat{\cal M}_0^a, \hat{\cal M}_0^b
{\Big]}_{+} \rangle M^b, \label{55}
\end{equation}
and
\begin{equation}
\Big[ \hat{\cal M}_0^a , \hat{\cal M}_0^b {\Big]}_{+} \equiv
\hat{\cal M}_0^a \hat{\cal M}_0^b + \hat{\cal M}_0^b \hat{\cal
M}_0^a. \label{56}
\end{equation}
The correlation function $G^{ab} \equiv \langle \Big[ \hat{\cal
M}_0^a , \hat{\cal M}_0^b {\Big]}_{+} \rangle$ is the sum of spin
correlation functions,
\begin{equation}
G^{ab} = 2{\gamma}^2 \sum_{i} \sum_{j \neq i} \langle \hat{S}_i^a
\hat{S}_j^b \rangle, \label{57}
\end{equation}
excluding the self-interaction of spins. For the standard ansatz
given in Eq.(\ref{52}), $\vec{\bf D}=0$ and the magnetization
equation does not change.

If the spin-spin interactions are turned on at $t=t_0$, so that
$\hat{\rho}(t_0)=\hat{\rho}_0(t_0)$, then, integrating both sides of
Eq.(\ref{14}), we find
\begin{equation}
\hat{\rho}_{\lambda}(t) = \hat{\rho}_0(t_0) +
\frac{\lambda}{i{\hbar}} \int_{t_0}^{t} d{\tau} \Big[ \hat{\cal
H}_{\lambda}(\tau), \hat{\rho}_{\lambda}(\tau) {\Big]}_{-}.
\label{58}
\end{equation}
Substituting Eq.(\ref{58}) into the definition of $G^{ab}$, yields
the equation
\[
G^{ab}(t) = G_0^{ab}(t_0)
\]
\begin{equation}
+ \frac{1}{\gamma} \int_{t_0}^t d{\tau} J^{cd}({\tau}) \Big(
{\varepsilon}^{ace} G^{ebd}({\tau}) + {\varepsilon}^{bce}
G^{aed}({\tau}) \Big), \label{59}
\end{equation}
where
\begin{equation}
G_0^{ab} \equiv \langle \Big[ \hat{\cal M}_0^a , \hat{\cal M}_0^b
{\Big]}_{+} {\rangle}_{0}, \label{60}
\end{equation}
which relates $G^{ab}$ to the third order correlation function, i.e.
the correlation function of the product of three magnetic moment
operators,
\begin{equation}
G^{abc} \equiv \langle \Big[ \Big[ \hat{\cal M}_0^a , \hat{\cal
M}_0^b {\Big]}_{+} , \hat{\cal M}_0^c {\Big]}_{+} \rangle.
\label{61}
\end{equation}
The correlation function $G^{abc}$, in turn, is related to the
fourth order correlation function and etc., and we have therefore an
infinite number of coupled equations for spin correlation functions.
For any practical calculation this infinite hierarchy has to be
truncated. That then defines the approximation scheme which may be
considered on the basis of the physical requirements for the system.
The approximation scheme will depend on the physical properties such
as density and on the strength of the interactions.

If the Hamiltonian $\hat{\cal H}_{I}$ is a small perturbation to the
original $\hat{\cal H}_0$, we can solve Eq.(\ref{58})
perturbatively. In the lowest, zeroth order in $\lambda$, we replace
$\hat{\rho}_{\lambda}(t)$ by $\hat{\rho}_0(t_0)$, so that $G^{ab}
\approx G_0^{ab}(t_0)$. We choose the initial value for $G^{ab}$ as
\begin{equation}
\sum_{i} \sum_{j \neq i} \langle \hat{S}_i^a \hat{S}_j^b {\rangle}_0
= I^{ab} \label{62}
\end{equation}
with $I^{xx}=I^{yy}=0$, $I^{zz}=I$ and $I^{ab}=0$ for $a \neq b$. We
also define again the $z$-direction as the direction of the
effective magnetic field that is chosen uniform and static. Then the
$\vec{\bf D}$-term becomes, in component form,
\begin{eqnarray}
D_x  & = & 2{\lambda} {\gamma} I M_y M_z,\\
D_y  & = & -2{\lambda} {\gamma} I M_x M_z, \\
D_z  & = &  0. \end{eqnarray}
producing two effects in the magnetization equation: the
magnetization is now precessing with respect to $(H_z - 2{\lambda}I
M_z)$, its z-component remaining constant in time, $(d/{dt})M_z=0$,
and the frequency of the precession is
\begin{equation}
\overline{\omega}_{0} \equiv {\omega}_0 \Big( 1- {\lambda}
\frac{2IM_z}{H_z} \Big). \label{64}
\end{equation}
Therefore, in the lowest order of perturbations, when the $\vec{\bf
D}$ is linear in ${\lambda}$, it shifts the direction and the
frequency of the precessional motion without introducing damping
effects. To find a role of the higher powers of ${\lambda}$ in
$\vec{\bf D}$ and to determine how they affect the magnetization
equation, a truncation of the chain of spin correlation equations is
needed. This would require a consistent perturbation approach to the
hierarchy of the coupled equations for the correlation functions.

\section{Conclusion}

We have derived a general form of magnetization equation for a
system of spins precessing in an effective magnetic field without
specifying the internal interactions. It can be applied in the study
of magnetization dynamics of any type, including non-equilibrium and
nonlinear effects, provided the interaction of individual spins with
each other and with other degrees of freedom of the system is
specified.

The $\vec{\bf D}$-term in the magnetization equation has been
obtained without using any approximation scheme. It is exact,
accumulates all effects of the internal interactions on the
magnetization precessional motion and can be a starting point for
practical calculations. For the spin-spin interactions, it is
determined by the spin correlation functions, which fulfil an
infinite chain of equations. A further analysis of the $\vec{\bf
D}$-term requires an approximation scheme to truncate the chain in a
consistent approach to higher order calculations.

In our talk, we have considered a specific type of the spin-spin
interactions, which do not contribute to the algebra of magnetic
moment operators. However, if spin-spin interactions depend
explicitly on the external field, the form of the algebra can
change. In this case, the total magnetic moment operator becomes
nonlinear in $\hat{\cal M}_0$, and this results in the magnetization
algebra with an infinite chain of commutation relations. The chain
has to be truncated in a way consistent with the truncation of the
chain of equations for the spin correlation functions in the same
approximation scheme.

Numerical computations along the lines developed in \cite{mars} can 
provide a further insight into the problem. This talk is based on the 
work \cite{skkm} where an extended list of references can be found.

\end{document}